\documentclass[journal]{IEEEtran}
\usepackage{amsmath,amssymb,amsfonts}
\usepackage{algorithmic}
\usepackage{amssymb}
\usepackage{algorithm,float}   
\usepackage{graphicx}
\usepackage{tabularx}
\usepackage{textcomp}
\usepackage{xcolor}
\usepackage{color}
\usepackage{stfloats}
\usepackage{enumerate}
\usepackage{multirow}
\usepackage{float}
\usepackage{bm}
\usepackage{ragged2e}
\usepackage{mathtools}
\usepackage[hidelinks]{hyperref}
\usepackage{float}

\newcommand\subparagraph{%
  \@startsection{subparagraph}{5}
  {\parindent}
  {3.25ex \@plus 1ex \@minus .2ex}
  {-1em}
  {\normalfont\normalsize\bfseries}}
\makeatother
\usepackage{titlesec}
\titleformat{\subsubsection}[runin]
        {\it}
        {\arabic{subsubsection})}
        {0.5em}
        {}
        [:]
\let\subparagraph\relax 
\titlespacing\section{0pt}{8pt}{4pt}
\titlespacing\subsection{0pt}{4pt}{2pt}


\usepackage{cite}

\usepackage{etoolbox}
\let\mybibitem\bibitem
\renewcommand{\bibitem}[1]{%
\ifstrequal{#1}{qi2013comparison}{\color{black}\mybibitem{#1}}
{\ifstrequal{#1}{caISO2011California}{\color{black}\mybibitem{#1}}
{\ifstrequal{#1}{martin2014synchrophasor}{\color{black}\mybibitem{#1}}
{\ifstrequal{#1}{pourramezan2017design}{\color{black}\mybibitem{#1}}
{\ifstrequal{#1}{mahapatra2017online}{\color{black}\mybibitem{#1}}
{\ifstrequal{#1}{mahapatra2016bad}{\color{black}\mybibitem{#1}}
{\ifstrequal{#1}{yang2020bad}{\color{black}\mybibitem{#1}}
{\ifstrequal{#1}{becejac2019impact}{\color{black}\mybibitem{#1}}
{\ifstrequal{#1}{idehen2019data}{\color{black}\mybibitem{#1}}
{\ifstrequal{#1}{idehen2016emulation}{\color{black}\mybibitem{#1}}
{\ifstrequal{#1}{zhang2014synchrophasor}{\color{black}\mybibitem{#1}}
{\ifstrequal{#1}{PMUstandard}{\color{black}\mybibitem{#1}}
{\ifstrequal{#1}{seyedi2019irregularity}{\color{black}\mybibitem{#1}}
{\ifstrequal{#1}{han2011data}{\color{black}\mybibitem{#1}}
{\color{black}\mybibitem{#1}}}}}}}}}}}}}}}%
}

\ifCLASSINFOpdf
\else
\fi
\begin{document}
%
\title{Cost-Effective Bad Synchrophasor Data Detection Based on Unsupervised Time Series Data Analytics
}
%
%
%

\author{Lipeng Zhu, \IEEEmembership{Member, IEEE,}
        and David J. Hill, \IEEEmembership{Life Fellow, IEEE}
        
\thanks{This work was supported in part by the Research Grants Council of Hong Kong Special Administrative Region under the Theme-based Research Scheme through Project No. T23-701/14-N and in part by The University of Hong Kong Research Committee Post-doctoral Fellow Scheme.}
\thanks{L. Zhu and D. J. Hill are with the Department of Electrical and Electronic Engineering, The University of Hong Kong, Hong Kong, China. (e-mail: zhulpwhu@126.com; dhill@eee.hku.hk)}


}

\maketitle

\begin{abstract}
In modern smart grids deployed with various advanced sensors, e.g., phasor measurement units (PMUs), bad (anomalous) measurements are always inevitable in practice. Considering the imperative need for filtering out potential bad data, {\color{black} this paper develops a novel online bad PMU data detection (BPDD) approach for regional phasor data concentrators (PDCs) by sufficiently exploring spatial-temporal correlations. With no need for costly data labeling or iterative learning, it performs model-free, label-free, and non-iterative BPDD in power grids from a new data-driven perspective of spatial-temporal nearest neighbor (STNN) discovery.} Specifically, spatial-temporally correlated regional measurements acquired by PMUs are first gathered as a spatial-temporal time series (TS) profile. Afterwards, TS subsequences contaminated with bad PMU data are identified by characterizing anomalous STNNs. To make the whole approach competent in processing online streaming PMU data, an efficient strategy for accelerating STNN discovery is carefully designed. {\color{black} Different from existing data-driven BPDD solutions requiring either costly offline dataset preparation/training or computationally intensive online optimization, it can be implemented in a highly cost-effective way, thereby being more applicable and scalable in practical contexts.} Numerical test results on the Nordic test system and the realistic China Southern Power Grid demonstrate the reliability, efficiency and scalability of the proposed approach in practical online monitoring.
 
\end{abstract}

\begin{IEEEkeywords}
Bad data detection, data analytics, spatial-temporal correlation, synchrophasor measurements, time series.
\end{IEEEkeywords}

%
\IEEEpeerreviewmaketitle

{\color{black}
\section*{Nomenclature}
\addcontentsline{toc}{section}{Nomenclature}
\begin{IEEEdescription}[\IEEEusemathlabelsep\IEEEsetlabelwidth{$V_1,V_2,$}]
\item[ANN] Artificial neural network.
\item[BPDD] Bad PMU data detection.
\item[CSG] China Southern Power Grid.
\item[DFT] Discrete Fourier transform.
\item[IoT] Internet of things.
\item[LOF] Local outlier factor.
\item[ML] Machine learning.
\item[OTW] Observation time window.
\item[PCA] Principal component analysis.
\item[PDC] Phasor data concentrators.
\item[PMU] Phasor measurement unit.
\item[SC] Spectral clustering.
\item[SE] State estimation.
\item[STNN] Spatial-temporal nearest neighbor.
\item[TS] Time series.
\item[WAMS] Wide-area measurement system.
\item[WT] Wavelet transform.
\item[$\bm A$] Ideal matrix of regional PMU measurements.
\item[$Acc$] Accuracy of BPDD.
\item[$d_{u,v}$] Distance between $\bm x'_{u,m}$ and $\bm x'_{v,m}$.
\item[$\bm d_u$] Vectorzied distances of $\bm x'_{u,m}$ to all the TS subsequences in $x '$.
\item[$\bm E$] Error matrix of regional PMU measurements.
\item[$Fal$] False alarm rate of BPDD.
\item[$\mathcal F(*)$] DFT manipulation.
\item[$\mathcal F^{-1}(*)$] Inverse DFT manipulation.
\item[$K$] Coefficient of the BPDD threshold $\xi$.
\item[$ m$] Length of TS subsequences for STNN characterization.
\item[$Mis$] Misdetection rate of BPDD.
\item[$M_{u,v}$] Dot product of $\bm x'_{u,m}$ and $\bm x'_{v,m}$.
\item[$N$] Number of elements (data points) in $\bm X$.
\item[$ n$] Number of data points within an OTW.
\item[$n_{\text{all}}$] Total number of instances for BPDD test.
\item[$ n_b$] Number of buses in a regional power grid.
\item[$n_{ta}$] Number of Truly detected anomalous instances.
\item[$n_{fn}$] Number of falsely dismissed anomalous instances.
\item[$n_{fa}$] Number of falsely alarmed normal instances.
\item[$Pre$] Precision of BPDD.
\item[$\bm p _{NN}$] Profile of all the TS subsequences' distances to their most similar TS subsequences (nearest neighbors) in $x '$, i.e., STNN profile.
\item[$\bm Q_{u}$] Inverse DFT result obtained from the dot products of $\bm X'_p$ and $\bm Y'_u$.
\item[$T$] Length (duration) of a given OTW.
\item[$\bm X$] Practical matrix of regional PMU measurements.
\item[$\bm X'_p$] DFT result of $\bm x'_p$.
\item[$\bm x_i$] TS of PMU measurements acquired from bus $i$.
\item[$\bm x '$] Concatenated TS consisting of all the elements in $\bm X$.
\item[$\bm x'_p$] Padded counterpart of $\bm x '$.
\item[$\bm x'_{u,m}$] $u$th TS subsequence extracted from $x '$.
\item[$\bm x'_{v,m}$] $v$th TS subsequence extracted from $x '$.
\item[$\bm Y'_u$] DFT result of $\bm y'_u$.
\item[$\bm y'_{u}$] Synthetic TS composed of the mirrored counterpart of $\bm x'_{u,m}$ and padded zero values.
\item[$\mu _u$] Mean value of  $\bm x'_{u,m}$.
\item[$\mu _v$] Mean value of  $\bm x'_{v,m}$.
\item[$\mu (\bm p _{NN})$] Mean value of  $\bm p _{NN}$.
\item[$\xi$] Threshold for STNN-based BPDD.
\item[$\sigma _u$] Standard deviation of  $\bm x'_{u,m}$.
\item[$\sigma _v$] Standard deviation of  $\bm x'_{v,m}$.
\item[$\sigma (\bm p _{NN})$] Standard deviation of  $\bm p _{NN}$.
\item[$\Delta t$] Sampling interval of PMU measurements.
\end{IEEEdescription}
}

\section{Introduction}
With the rapid development of smart sensing and Internet of things (IoT) technologies, advanced wide-area measurement systems (WAMS) have been increasingly deployed in modern smart grids\cite{bedi2018review, james2019synchrophasor}. By acquiring high-resolution measurement data using phasor measurement units (PMUs) in a synchronized manner, the WAMS significantly enhance power grids' capability  in situational awareness during online monitoring. In this circumstance, the PMU data quality acts as the cornerstone of many WAMS-based advanced applications, such as online dynamic stability assessment \cite{xu2016assessing}, wide-area event detection \cite{ge2015power_event}, and wide-area stability control \cite{kamwa2012compliance}. 

However, due to the inevitability of sensing errors and WAMS component malfunctions, bad (anomalous) PMU data are widely witnessed in practice. {\color{black} According to statistical analysis by power system practitioners in China, about 10\%$\sim$30\% of PMU measurements in China's grids are contaminated with bad data \cite{qi2013comparison, yang2020bad}. In North America, the bad PMU data ratio may be somewhat lower, yet still being far from satistactory. As reported by California Independent System Operator, its bad PMU data proportion is generally 10\%$\sim$17\% \cite{caISO2011California, wu2017online}.} Moreover, as PMU measurements acquired during online monitoring come into {\color{black} phasor data concentrators (PDCs) and upper control rooms} as data streams, bad data should be timely filtered out to avoid undesirable data accumulation. Therefore, it is imperative to develop reliable and efficient bad PMU data detection (BPDD) schemes for practical grids.

In terms of BPDD, the research community has made tremendous efforts to tackle this problem during power system online monitoring. Conventionally, state estimation (SE) related approaches are widely adopted to perform online BPDD. In \cite{korres2011state}, with the help of a non-linear weighted least squares state estimator, normalized residual tests are performed to identify bad data. An augmented state vector approach is proposed for SE in \cite{ghiocel2014phasor}, which is able to both detect bad PMU data and improve the data quality. Recently, by classifying suspicious data into small groups and implementing largest normalized residual tests in parallel, a highly efficient SE based BPDD method is reported in \cite{lin2018highly}. Based on quadratic prediction and Kalman filtering, an algorithm is proposed in \cite{jones2014methodology} to preprocess bad PMU data before performing SE. {\color{black} In addition, some recent PMU-based applications have taken the problem of BPDD into account to improve their applicability. For example, focusing on identifying multiple power line outages in the presence of bad PMU data, a systematic framework is developed in \cite{li2016location}, which helps correct bad measurements simultaneously. With the dual target of enhancing system observability and bad data detection, a unified PMU placement scheme for wide-area SE is proposed in \cite{gou2014unified}.} While these methods have exhibited their strength in coping with BPDD in their case studies, their reliability could be impaired in practice due to their heavy reliance on system topological information and model parameters.    

In recent years, a handful of inspiring data-driven efforts have shown high potential in fulfilling the BPDD task. In \cite{ gao2016missing}, based on the low-rank property of spatial-temporal measurement matrices, missing PMU data are detected and recovered by solving a low-rank matrix completion problem. Similarly, the low-rank property of the Hankel structure is exploited to identify and correct bad PMU data in \cite{hao2018modelless}. Recently, on the basis of the intrinsic spatial-temporal correlations between multiple PMU channels, a density based clustering method called local outlier factor (LOF) analysis is introduced in \cite{wu2016online, wu2017online} for BPDD. Essentially, these methods exploit  power systems' {\color{black} inherent spatial-temporal properties/correlations reflected in regional PMU measurements \cite{wu2017online, seyedi2019irregularity}} to detect potential anomalies. Compared with the afore-mentioned model-based solutions, these model-free alternatives would achieve more reliable BPDD in the presence of inaccurate topology information or parameter errors. Nevertheless, they have their own limitations. As the low-rank based approaches in \cite{gao2016missing, hao2018modelless} involve complicated optimization procedures to solve the BPDD problem, their implementations are likely to be computationally expensive in practical onling monitoring. While the LOF-based method \cite{wu2016online, wu2017online} carries out  BPDD at a high speed, its online reliability depends on the preparation of a high-quality historical PMU database (with no bad data), which requires tough class labeling efforts to screen out all the anomalous historical data. {\color{black} Recently, a
principal component analysis (PCA) based BPDD method is reported in \cite{mahapatra2016bad}. With data points statistically processed, however, it does not sufficiently explore the temporal correlations between sequential data points, which may limit its performance. In \cite{mahapatra2017online}, this PCA technique is further combined with the artificial neural network (ANN) based machine learning (ML) method for enhanced BPDD. Similarly, the spectral clustering (SC) algorithm is integrated with the decision tree technique to develop a ML scheme for BPDD \cite{yang2020bad}. These two ML-based solutions require a labelled dataset for offline training, and it could be challenging for them to generalize to unforeseen bad data scenarios.}     

Taking the above research gaps into account, {\color{black} this paper develops an efficient model-free BPDD approach for regional PDCs via unsupervised time series (TS) data analytics.} In particular, with sequential PMU measurements in a specific region integrated as a spatial-temporal TS profile, the BPDD problem is first converted to spatial-temporal anomaly detection from TS. Then, sequential BPDD is efficiently performed by identifying anomalous TS subsequences which remain far away from their spatial-temporal nearest neighbors (STNN). {\color{black} As their STNN values are expected to be significantly different from normal subsequences, BPDD can be reliably carried out via adaptive STNN threshold-based checking rules.}  

{\color{black}The reason why unsupervised data analytics is considered in this paper lies in that it can circumvent the need for domain expertise to perform costly data labeling; instead, by implementing automatic feature discovery, it can enhance the BPDD scheme's applicability in practice. For the sake of sufficiently exploiting spatial-temporal correlations to perform BPDD, TS data analytics is further integrated into it for sequential feature characterization from system dynamics. Compared with representative data-driven BPDD solutions in the literature\cite{gao2016missing, hao2018modelless, wu2016online, wu2017online, mahapatra2017online, yang2020bad}, the BPDD approach proposed in this paper has the following advantages:}
\begin{itemize}
\item
The proposed approach does not involve time-consuming offline training that is widely witnessed in conventional big data analytics solutions. In addition, {\color{black} unlike existing data-driven BPDD methods requiring costly data processing for offline clean/labelled dataset preparation \cite{wu2016online, wu2017online, mahapatra2017online, yang2020bad}, it has no offline computational cost. 
\item
By characterizing distinct STNN profiles, the BPDD approach can precisely detect various typical bad data with apparently indistinguishable profiles. With no need for computationally intensive online optimization \cite{gao2016missing, hao2018modelless}, it leverages a fast STNN discovery technique to perform BPDD in a non-iterative manner. As a result, very few online computational efforts are needed in practice.}
\item
During practical application, after the completion of online data acquisition, it can simply and efficiently work in a "plug-and-play" fashion. {\color{black} Specifically, with no requirement on performing offline data preparation/training in advance, it can immediately work online after its functionality is switched on.}
\end{itemize}

{\color{black} Owing to the first two salient features w.r.t. offline/online computational costs, the BPDD approach can be realized in a computationally cost-effective way in practical power grids. Due to the third attractive feature, this method is capable of addressing online PMU data streams in a highly efficient manner. As will be demonstrated by experimental tests in the sequel,} the proposed approach has strong applicability and scalability in practical contexts. The remainder of the paper is structured as follows. Section \ref{section 2} describes the BPDD problem and formulates it as detecting spatial-temporal anomalies. Section \ref{section 3} presents the model-free sequential BPDD approach in detail. In Sections \ref{section 4} and \ref{section 5}, numerical tests are extensively carried out on the Nordic test system and the real-world CSG to comprehensively test its BPDD performances. {\color{black} Section \ref{section 6} provides further discussions on the proposed approach.} Finally, Section \ref{section 7} concludes the paper.

\section{Bad PMU Data: Spatial-Temporal Anomaly} \label{section 2}
\subsection{Problem Description}
For a certain region in a given power grid, suppose $n_b$ PMUs are installed at $n_b$ neighboring buses for online monitoring. {\color{black} When these buses are located in a small region, they are expected to present relatively strong electrical couplings.} Given an observation time window (OTW) of length $T$, PMU measurements are sequentially acquired from individual buses with a sampling interval of $\Delta t$. Sequential PMU data of a certain type of electrical quantities, e.g., voltage magnitude, are gathered as a spatial-temporal measurement matrix consisting of $n_b$ TS:
\begin{equation}
{\bm X } = \begin{bmatrix}
\bm x _1 \\
\bm x _2 \\
\vdots \\
\bm x _{n_b}
\end{bmatrix}
= \begin{bmatrix}
x_{11}& x_{12} & \cdots & x_{1,n} \\
x_{21}& x_{22} & \cdots & x_{2,n} \\
\vdots & \vdots & \vdots & \vdots \\
x_{n_b, 1}& x_{n_b, 2} & \cdots & x_{n_b, n}
\end{bmatrix}
\label{X}
\end{equation} 
where $\bm x _i = \{x_{i,1}, x_{i,1}, ..., x_{i,n}\}$ is the TS of PMU measurements obtained from bus $i$ ($1\leq i \leq n_b$), and $n = T/\Delta t$ is the number of data points in $\bm x _i$. In fact, $\bm X$ can be decomposed into two independent parts:
\begin{equation}
{\bm X } = \bm A + \bm E 
= \begin{bmatrix}
\bm \alpha _1 \\
\bm \alpha _2 \\
\vdots \\
\bm \alpha _{n_b}
\end{bmatrix}
+ \begin{bmatrix}
\bm \epsilon _1 \\
\bm \epsilon _2 \\
\vdots \\
\bm \epsilon _{n_b}
\end{bmatrix}
\label{A+E}
\end{equation}
where $\bm A = [\bm \alpha _1, \bm \alpha _2, ..., \bm \alpha _{n_b}]^T$ and $\bm E = [\bm \epsilon _1, \bm \epsilon _2, ..., \bm \epsilon _{n_b}]^T$ represent the ideally measured states of the power grid and the measurement errors, respectively. 

In essence, due to the inherent networked electrical couplings between individual buses, neighboring buses generally have similar dynamic behaviors in both normal quasi-steady states and dynamic processes caused by transient events. Hence, relatively strong spatial-temporal correlations dwell in the $n_b$ TS of $\bm A$, i.e., the sequences  $\bm \alpha _1, \bm \alpha _2, ..., \bm \alpha _{n_b}$ share significant similarities with each other. $\bm X$ is expected to inherit such spatial-temporal correlations from $\bm A$ provided that the measurement errors in $\bm E$ are trivial. However, if the measurement errors become too large, the intrinsic spatial-temporal correlations cannot be sufficiently captured by the matrix $\bm X$. Based on the definition in \cite{wu2016online}, such PMU data with large measurement errors are called bad (low-quality) PMU data. {\color{black} Note that the extreme case of complete PMU data loss can be also considered as bad data by filling the lost PMU measurements with zero values.} In fact, as the spatial-temporal patterns characterized by bad PMU data are significantly different from the original spatial-temporal characteristics of the actual system dynamics, the corresponding sequential bad data are deemed as spatial-temporal anomalies in $\bm X$. In this regard, the fundamental task of detecting bad PMU data can be converted to identifying spatial-temporal anomalies in $\bm X$. 

{\color{black} It should be clarified that this paper focuses on online detection of typical bad data that cannot be directly handled by conventional signal processing tools or engineering experience based rules \cite{martin2014synchrophasor} deployed in local PMUs or single substations. Such data range from data spikes to repeated (un-changed) data and false data injection \cite{idehen2016emulation, wu2017online, becejac2019impact}, etc. While outliers like data spikes can be easily identified in stationary operating conditions by checking if sudden changes occur, this could fail in the presence of transient events which induce similar sudden changes as well. Besides, though with a limited influence, repeated small segments in quasi-steady states may not be successfully detected by conventional signal filtering or calibrating mechanisms. During transient periods, if PMU measurements keep unchanged from the pre-fault stage to the post-fault stage, these outliers may be difficult to be detected due to the similar profiles w.r.t. switching events or sudden load changes. As for false data injection, since deceptive data may be modified by cyber attackers to any values, it could be very tough to distinguish those malicious data highly resembling actual system states. To tackle these challenging BPDD issues, instead of relying on isolated information obtained from single substations, this paper will exploit the above-mentioned spatial-temporal correlations in regional PMU measurements for BPDD. With its regional scope, the proposed BPDD method can be deployed in regional PDCs.    

Note that, although communication problems such as network latency/corruption and packet loss also induce bad data issues in the form of data dropout or temporary data unavailability, these anomalies do not need special BPDD efforts due to the default use of NaN (not a number) or zero values to flag such visible bad data \cite{PMUstandard, idehen2019data}. In addition, while bad data in the form of high sensing noises are studied in some existing BPDD schemes \cite{wu2017online, becejac2019impact}, they are not taken into main account in this paper, because they can be directly identified by local signal processing tools.}         

\subsection{Illustrative Example}
Taking data spikes along with similar event data for example, how to exploit spatial-temporal correlations to detect bad PMU data is illustrated using real-world PMU data acquired from three adjacent buses in CSG. {\color{black} The raw voltage profile of bus 1 is depicted in Fig.~\ref{Fig1_bad_data_illustration}(a). As can be seen, it has four sharp drops to zero values and two moderate data spikes. Based on engineering experiences, one can directly identify that the four sharp drops are bad data points caused by data loss or dropout. However, it may be difficult even for experienced practitioners to ascertain whether the remaining two data spikes involve natural voltage sags caused by physical disturbances or bad measurements due to PMU/WAMS malfunction. In fact, as physical disturbances/events may not be a priori known during online monitoring, how to distinguish event-related sags from data quality-induced spikes merely based on local bus measurements is a nontrivial problem \cite{wu2017online}.

\begin{figure}[t] 
	\color{black}
	\centering
	\includegraphics[width=1.0\linewidth]{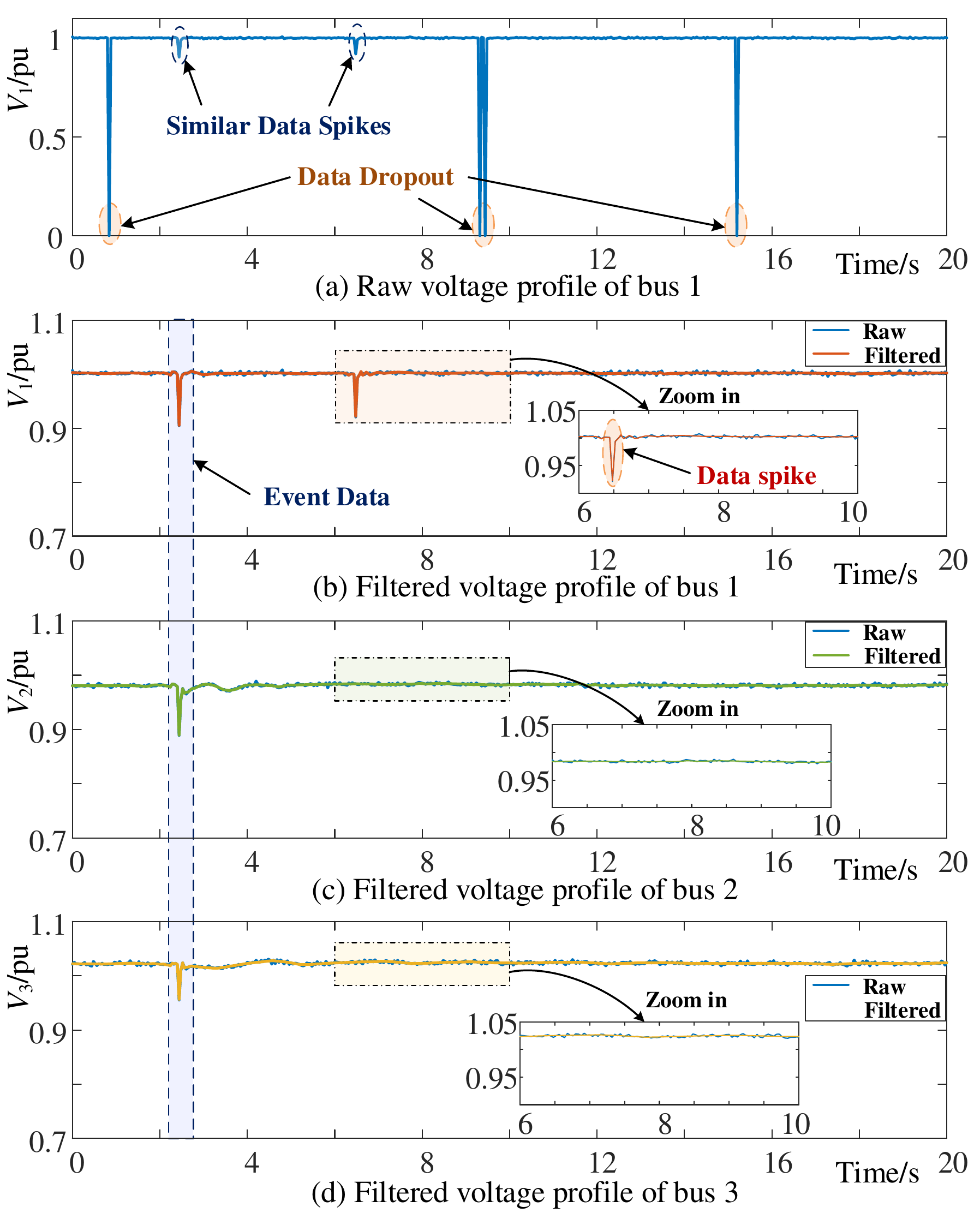}
	\caption{Illustration of bad PMU data in a real-world power grid.}
	\label{Fig1_bad_data_illustration}
\end{figure}

With the four abnormal zero-value data points corrected via linear interpolation, the partially fixed voltage profile of bus 1 is illustrated in Fig.~\ref{Fig1_bad_data_illustration}(b). Taking wavelet transform (WT) as a standard denoising tool, potential measurement noises in the profile are filtered out. Similarly, WT-based denoising is carried out on the voltage profiles of buses 2 and 3, as presented in Fig.~\ref{Fig1_bad_data_illustration}(c)$\sim$(d). By comparing the differences of filtered voltage profiles in Fig.~\ref{Fig1_bad_data_illustration}(b)$\sim$(d), one can easily figure out the first data spike at 2.48 s is likely to be caused by a physical event, since consistent data spikes are also observed in the other two buses. As this belong to normal system behaviors, it cannot be considered as anomalous measurements. For the second data spike at 6.52 s, it is deduced to be bad data, because only bus 1 undergoes the sudden dip, whereas buses 2 and 3 present relatively smooth evolution.} 

In fact, this simple BPDD example implicitly exploits the knowledge about spatial-temporal correlations between the three buses, i.e., adjacent bus voltages are expected to have similar dynamic voltage profiles. It shows the high potential taking advantage of spatial-temporal correlations for BPDD in the presence of distracting measurements like event data. As shown in Fig.~\ref{Fig1_bad_data_illustration}, without bad data, contemporary event data of adjacent buses still share similar evolution trends with each other, i.e., they still sufficiently capture the inherent spatial-temporal features in the physical system. In contrast, bad data are likely to break the original spatial-temporal patterns of the actual system dynamics. {\color{black} As will be shown below, these features can be exploited to formally develop a powerful BPDD approach from a new perspective of STNN discovery.}

\section{Sequential Bad PMU Data Detection} \label{section 3}
In the spatial-temporal measurement matrix $\bm X$, if it does not contain bad data, the $n_b$ TS will resemble each other to some degree. {\color{black} Under this circumstance, each subsequence extracted from the TS would have at least one neighbor staying relatively close to itself.} Otherwise, a TS subsequence including bad data would be a spatial-temporal anomaly that is significantly different from other subsequences, thus being very dissimilar to its nearest neighbor. Following these basic ideas, an  efficient sequential BPDD approach by characterizing STNN is developed in this paper. Its technical details are described in the following.  

\subsection{Profiling STNN for BPDD} 
With  all the $n_b$ TS concatenated one by one, the $n_b \times n$ measurement matrix $\bm X$ is reshaped into a vectorized TS:
\begin{equation}
{\bm x '} = [x'_1, x'_2, ..., x'_k, ..., x'_N],~\text{for}~N = n_b \times n
\label{x'}
\end{equation}
where $x'_k = x_{i,j}$ is the original entry in the $i$th row ($1\leq i \leq n_b$) and $j$th column ($1\leq j \leq n$) of $\bm X$, with $i = ceil(k/n)$ (rounding towards $+\infty$) and $j = k - (i - 1)*n$. By designating a length value $m$ ($3\leq m \leq N$), $(N - m + 1)$ subsequences of length $m$ are extracted from $\bm x'$:
\begin{equation}
\bm x'_{u,m} = \{x'_u, x'_{u+1},..., x'_{u+m-1}\},~\text{for}~1\leq u \leq N-m+1 
\label{subsequence}
\end{equation}

Given two subsequences $\bm x'_{u,m}$ and $\bm x'_{v,m}$ ($1\leq u,v \leq N-m+1$), their dissimilarity is measured by
\begin{equation}
d_{u,v} =  \sqrt{2m\left(1 - \frac{M_{u,v} - m \mu_u \mu_v}{m \sigma_u \sigma_v} \right)}  
\label{correlation_distance}
\end{equation}
where $M_{u,v} = \sum_{k=1}^{m}x_{u+k-1}x'_{v+k-1}$ is the dot product of $\bm x'_{u,m}$ and $\bm x'_{v,m}$, $\mu_u$ and $\mu_v$ are their mean values, and $\sigma_u$ and $\sigma_v$ are their standard deviations, respectively. Note that, as illustrated in \cite{mueen2010fast,zakaria2012clustering}, \eqref{correlation_distance} is equivalent to the normalized Euclidean distance. The reason why such an apparently more complex distance measure is adopted here is that it can achieve significant acceleration for vectorized STNN profile calculations (see Section III-B). 

For each subsequence in $\bm x'$, by calculating its distances to all the subsequences in $\bm x'$, a distance vector is obtained:
\begin{equation}
\bm d_u = \left[d_{u,1},d_{u,2},..., d_{u,N-m+1}\right]~(1\leq u \leq N-m+1)
\label{distance_vector}
\end{equation}
The minimum distance values are then collected from each distance vector (with $d_{u,u}$ deleted from $\bm d_u$ to avoid trivial minimum estimation) to form a vectorized profile:
\begin{equation}
\bm p_{NN} = \left[\min (\bm d_1), \min (\bm d_2), ..., \min (\bm d_{N-m+1})\right]
\label{NN_profile}
\end{equation}
As $\bm p_{NN}$ represents the collection of all the subsequences' distances to their  nearest neighbors in the spatial-temporal measurement matrix, it is thus called the STNN profile. By setting a threshold $\xi$, the subsequence $\bm x'_{u,m}$ would be identified as an anomaly with bad PMU data points if its STNN distance satisfies the following condition:
\begin{equation}
\bm p_{NN}(u) = \min (\bm d_u) > \xi
\label{criterion}
\end{equation}
The condition in \eqref{NN_profile} indicates that $\bm x'_{u,m}$ is significantly different from others, with its nearest neighbor being far away from it. Hence, bad PMU data are expected to exist in $\bm x'_{u,m}$. As this is an abnormal TS subsequence, it is also called TS discord in the TS data mining community \cite{yeh2016matrix}. In order to let this criterion be adaptive to statistical characteristics of different spatial-temporal profiles, the threshold for decision is automatically determined by
\begin{equation}
\xi = \mu (\bm p_{NN}) + K \sigma (\bm p_{NN})
\label{threshold}
\end{equation}   
where $\mu (\bm p_{NN})$ and $\sigma (\bm p_{NN})$ are the mean value and standard deviation of $\bm p_{NN}$, and $K$ is a coefficient controlling the criterion's sensitivity. Based on \eqref{threshold}, $\xi$ represents a certain anomalous level, which is similar to the $3\sigma$ rule in Gaussian distributions. Empirical tests show that setting the coefficient to $K = 6$ generally results in highly reliable BPDD.      

\subsection{Fast STNN Discovery} 
The derivation of STNN in \eqref{NN_profile} involves numerous distance calculations and comparisons, especially when the total number of subsequences in $\bm x'$ is very large. With a brute-force manner of calculating and comparing pair-wise distances one by one, the overall computational complexity of profiling STNN would be extremely high. If not treated properly, such a heavy computational burden will deteriorate the proposed approach's online performances in practice, where streaming PMU data need to be processed efficiently. To speed up online BPDD, a fast STNN discovery strategy is introduced in this paper based on a novel pairwise similarity search algorithm \cite{yeh2016matrix, zhu2016matrix}.

The key to accelerating the procedure of STNN discovery lies in improving the efficiency of pairwise distance calculation in \eqref{correlation_distance}. As the mean values and standard deviations in \eqref{correlation_distance} can be efficiently computed by existing commercial software such as \textsc{Matlab}, the main concern is how to quickly obtain the dot product $M_{u,v}$. In this paper, the convolution based discrete Fourier transform (DFT) and its inverse counterpart \cite{yeh2016matrix} are exploited to perform batch dot-product computations for $\bm d_u$ in a vectorized manner. Before the DFT manipulation, two synthetic sequences are first derived from $\bm x'$ and $\bm x'_{u,m}$ by padding zeros and re-ordering elements in a mirrored way:
\begin{equation}
\bm x'_p = [x'_1, x'_2, ..., x'_N, \underbrace{ 0, 0, \cdots, 0}_ {N~\text{zeros}}]
\label{x'_p}
\end{equation}
\begin{equation}
\bm y'_u = [x'_{u+m-1}, x'_{u+m-2}, ..., x'_u, \underbrace{ 0, 0, \cdots, 0}_ {(2N-m)~ \text{zeros}}]
\label{x_u_inverse}
\end{equation}

Then, DFT is performed on $\bm x'_p $ and $\bm y'_u$, which yields
\begin{equation}
\arraycolsep=0.5pt\def\arraystretch{1.2}
\left\{ {\begin{array}{lr}
\bm X'_p = \mathcal F (\bm x'_p ) = [X'_1, X'_2, ..., X'_{2N}]\\
\bm Y'_u = \mathcal F (\bm y'_u) = [Y'_{u1}, Y'_{u2}, ..., Y'_{u,2N}]
\end{array}} \right.\
\label{DFT}
\end{equation}   
where $\mathcal F(*)$ denotes DFT. Based on the DFT calculation results, inverse DFT is further carried out:
\begin{equation}
\bm Q_u = \mathcal F^{-1} (\bm X'_p \odot \bm Y'_u) = [Q_{u1}, Q_{u2}, ..., Q_{u,2N-1}]
\label{iDFT}
\end{equation} 
where $\mathcal F^{-1}(*)$ represents inverse DFT, and $\odot$ denotes the dot product of $\bm X'_p$ and $\bm Y'_u$. In fact, as demonstrated in \cite{yeh2016matrix}, all the pairwise dot products between $\bm x'_{u,m}$ and other subsequences are included in $\bm Q_u$, and they can be efficiently retrieved as 
\begin{equation}
M_{u,i} = Q_{u,m-1+i},~\text{for}~ 1\leq i \leq N-m+1
\label{M_ui}
\end{equation} 
  
As can be observed in \eqref{x'_p}-\eqref{M_ui}, by performing sequential dot products once and DFT based calculation three times, the original estimation of $(N-m+1)$ sequential dot-products can be quickly finished in a vectorized way. This would lead to a significant acceleration for searching STNNs. Moreover, based on the recursive relationship between successive dot products, further speed-up can be achieved. Given the dot product value $M_{u,v}$, one can easily estimate $M_{u+1,v+1}$ as 
\begin{equation}
M_{u+1,i+1} = M_{u,v} - x'_u x'_v + x'_{u+m} x'_{v+m}
\label{M_new}
\end{equation}
Based on the above preliminaries, the following strategy is adopted to efficiently compute dot products for all the subsequences: 1) $\{M_{11}, M_{12},..., M_{1,N-m+1}\}$ are quickly calculated using \eqref{x'_p}-\eqref{M_ui}; 2) the remaining dot products are recursively computed via \eqref{M_new}. With this simple strategy, the procedure of STNN discovery could be tens to hundreds of times faster than the brute force method.

\section{Simulation Test in Benchmark System} \label{section 4}
\begin{figure}[t]
	\centering
	\includegraphics[width=0.9\linewidth]{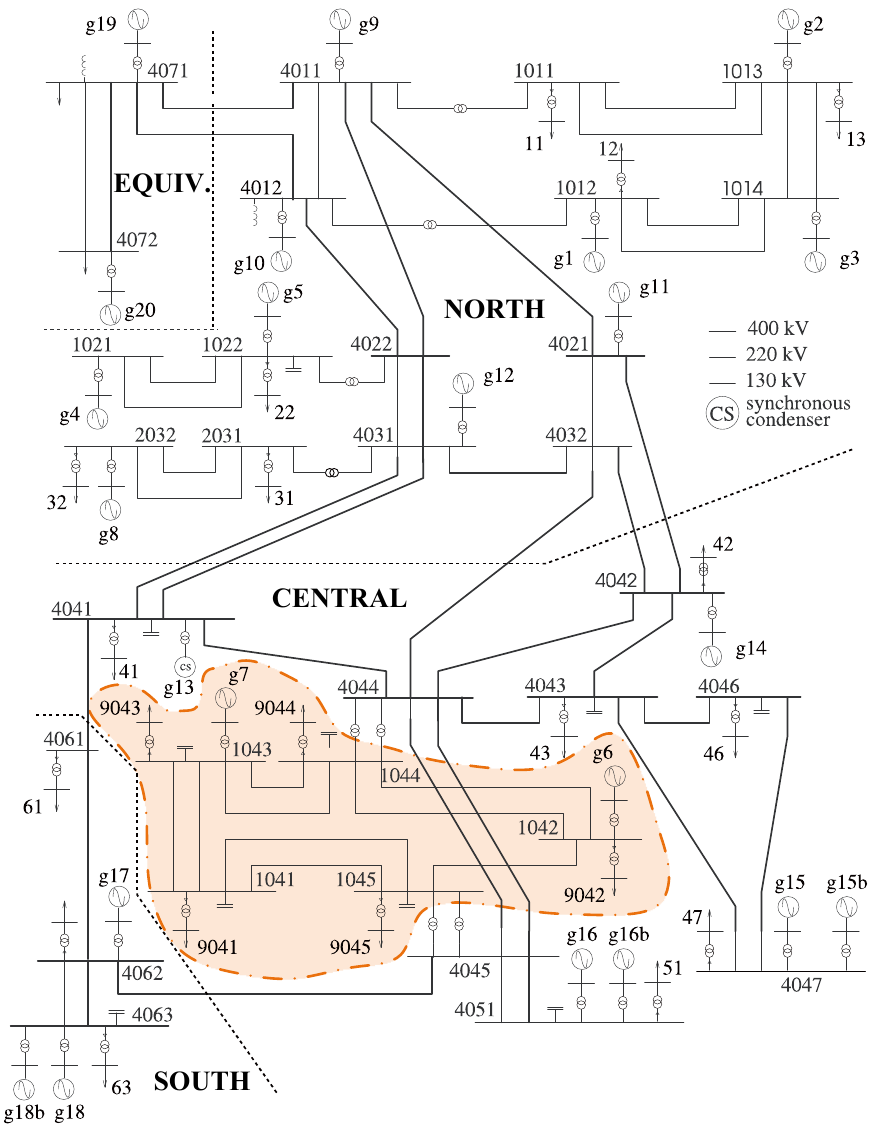}
	\caption{One-line diagram of the Nordic test system.}
	\label{Fig8_Nordic}
\end{figure}

The proposed approach was first tested on the Nordic test system to verify its efficacy. This is a benchmark system simplified from the actual Swedish and Nordic power grid \cite{van2013description, ospina2017implementation, zhu2017imbalance}. As shown in the shaded area of Fig.~\ref{Fig8_Nordic}, the five adjacent 130-kV buses in its receiving-end area, i.e., buses 1041$\sim$1045, were assumed to be configured with PMUs for online monitoring. 
Supposing that the system encountered transient events such as three-phase short circuits, time-domain simulations were conducted to simulate the acquisition of sequential PMU data during system dynamics. Specifically, PMU data were acquired with a sampling rate of 100 Hz. The length of the OTW was set to 5 s. With 500 data points in the OTW, the length of subsequences for STNN discovery was empirically specified as $1/10$ of the OTW, i.e., $m = 50$.

{\color{black}Taking into account typical operating conditions, topological changes and fault locations in the Nordic test system, 1600 dynamic cases subject to various contingencies were generated by time-domain simulations. Typical bad data, including data spikes, repeated values and false data injection, were randomly imposed onto the bus voltage profiles to simulate anomalies.} In each case, an OTW sliding with step $\Delta T = 1$ s was utilized to extract five windows of voltage related PMU data matrices from buses 1041$\sim$1045. By doing so, 8000 PMU measurement matrices were obtained for BPDD. 

During STNN discovery, all the TS from different buses were normalized by their respective steady-state measurements. To avoid trivial anomaly detection of adjacent subsequences with substantial overlaps, a sliding step with length of $m/10 = 5$ was set to detect spatial-temporal anomalies in the STNN profile (the same below). In particular, starting from the peak point with the largest STNN distance, if it satisfies the criterion in \eqref{criterion}, the corresponding subsequence is identified as a spatial-temporal anomaly. Then, the two neighboring data points with five steps from the peak point are examined using \eqref{criterion}. {\color{black} Before testing the overall BPDD performances, the potential of the proposed approach in BPDD is first demonstrated with typical examples below.} 
  
\subsection{BPDD against Data Spike}
\begin{figure*}[t] 
	\color{black}
	\centering
	\includegraphics[width=1.0\linewidth]{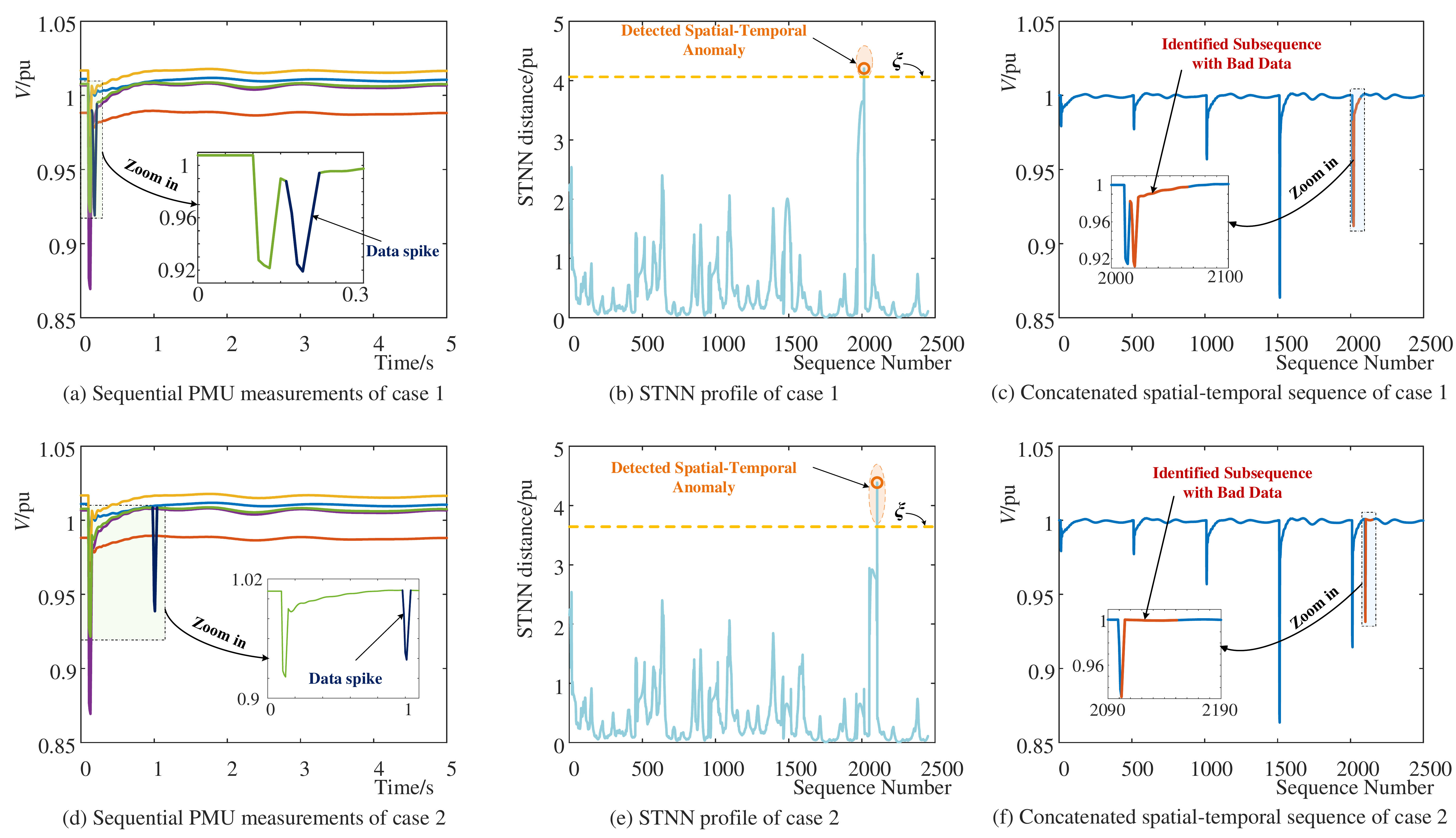}
	\caption{BPDD against data spikes. (case 1 $\rightarrow$ data spike in fault-on stage; case 2 $\rightarrow$ data spike in post-fault stage.)}
	\label{Fig2_scenario1}
\end{figure*}

{\color{black} As shown in Fig.~\ref{Fig2_scenario1}, two cases involving data spikes are chosen for illustration here. For the first case, a data spike is imposed onto system fault-on dynamics induced by a short-circuit event. As can be seen in Fig.~\ref{Fig2_scenario1}(a), two highly similar sudden voltage sags are observed in the fault-on stage. Without prior knowledge, it is difficult to confirm whether the first spike is anomalous (with a physical event arising at about 0.15 s) or the second one is an outlier (with a physical event occurring at about 0.1 s). As illustrated in Fig.~\ref{Fig2_scenario1}(b)$\sim$(c), the proposed approach is able to distinguish them clearly. It not only accurately recognizes the anomalous data spike (second spike), but also bypasses the event-related voltage dip (first spike).} Such a strong discriminability reveals that the inherent spatial-temporal correlations within the system indeed help to differentiate abnormal data spikes from its natural dynamics.  

{\color{black} For the second case in Fig.~\ref{Fig2_scenario1}(d), a data spike with a relative similarity to transient voltage sags occurs in the post-fault stage. Compared with the above case, this one is easier to be identified by observing the post-dip differences between actual system disturbances (with fluctuating dynamics) and anomalous data spikes (with relatively stationary states). It can be easy for experienced engineers to carry out this detection, yet the proposed approach with no resort to any domain expertise can also automatically find such differences for BPDD, as shown in Fig.~\ref{Fig2_scenario1}(f). This indicates that the proposed approach has a desirable capability in automatic feature discovery.}

\subsection{BPDD against Repeated Data} 
\begin{figure*}[t] 
    \color{black}
	\centering
	\includegraphics[width=1.0\linewidth]{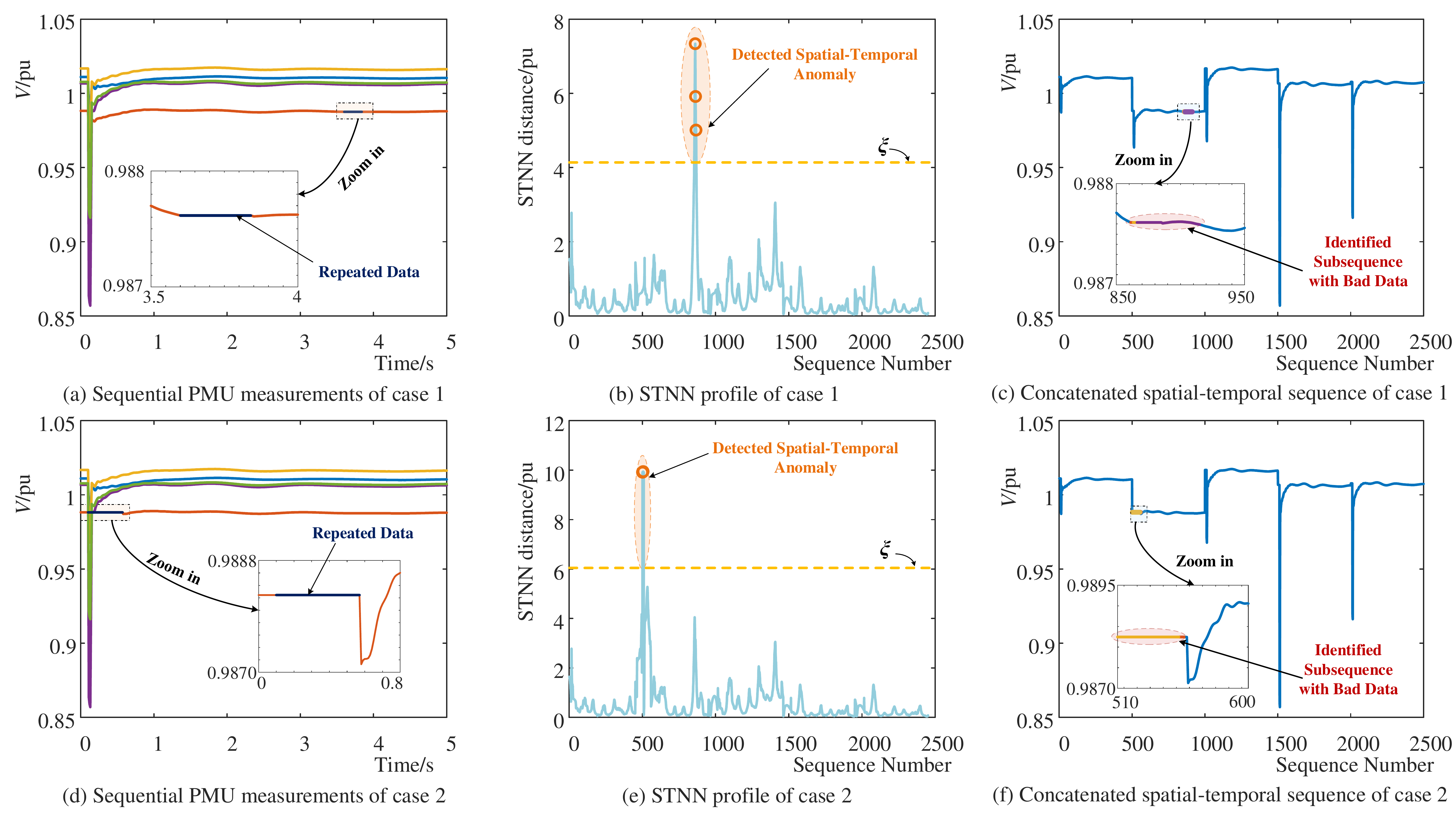}
	\caption{BPDD against repeated data. (case 1 $\rightarrow$ repeated data in quasi-steady state; case 2 $\rightarrow$ repeated data in transient process.)}
	\label{Fig3_scenario2}
\end{figure*}

{\color{black} How BPDD is implemented in the presence of repeated voltage values is exemplified with two cases in Fig.~\ref{Fig3_scenario2}. The first case involves un-updated data in the quasi-steady state. As depicted in Fig.~\ref{Fig3_scenario2}(a), the un-updated measurements seem to be indistinguishable from normal quasi-steady-state values even with a zoom-in view. With the help of the proposed approach, such un-updated data are successfully detected by checking the slight contemporary differences between the five buses.}  

{\color{black} As for the second case, repeated data last from the pre-fault stage to the post-fault stage, which makes the corresponding bus voltage profile miss fault-on system responses. As shown in Fig.~\ref{Fig3_scenario2}(d), such un-updated measurements are also difficult to be identified by observing a single bus voltage profile. In fact, the repeated data along with the subsequent sudden voltage dip resemble system responses caused by abrupt physical disturbances like sudden load changes or switching events. With special attention to spatial-temporal correlations among the five buses, the proposed approach correctly identifies these abnormally un-updated data. Note that this is achieved in the presence of much more fluctuating system transients,} which implies that false alarm is avoided on the event data.

{\color{black}
\subsection{BPDD against False Data Injection}
Considering possible cyber attacks at substations, two cases with false data injection are chosen for further illustration of BPDD, as presented in Fig.~\ref{Fig4_scenario3}. Assuming the cyber attacker has retrieved some historical PMU measurements from the system, deceptive data injection is carried out with realistic historical data rather than simple synthetic signals that can be easily detected by existing false data processing tools at individual substations. Specifically, a segment of historical oscillation signals is injected into post-fault system dynamics in the first case. Due to the intrinsic oscillatory system behavior after fault clearance, such deceptive oscillations cannot be directly identified even if manual comparisons with contemporary measurements from adjacent buses are made. In this highly deceptive scenario, the proposed BPDD approach still detects the occurrence of these falsely injected data accurately.

\begin{figure*}[t] 
	\color{black}
	\centering
	\includegraphics[width=1.0\linewidth]{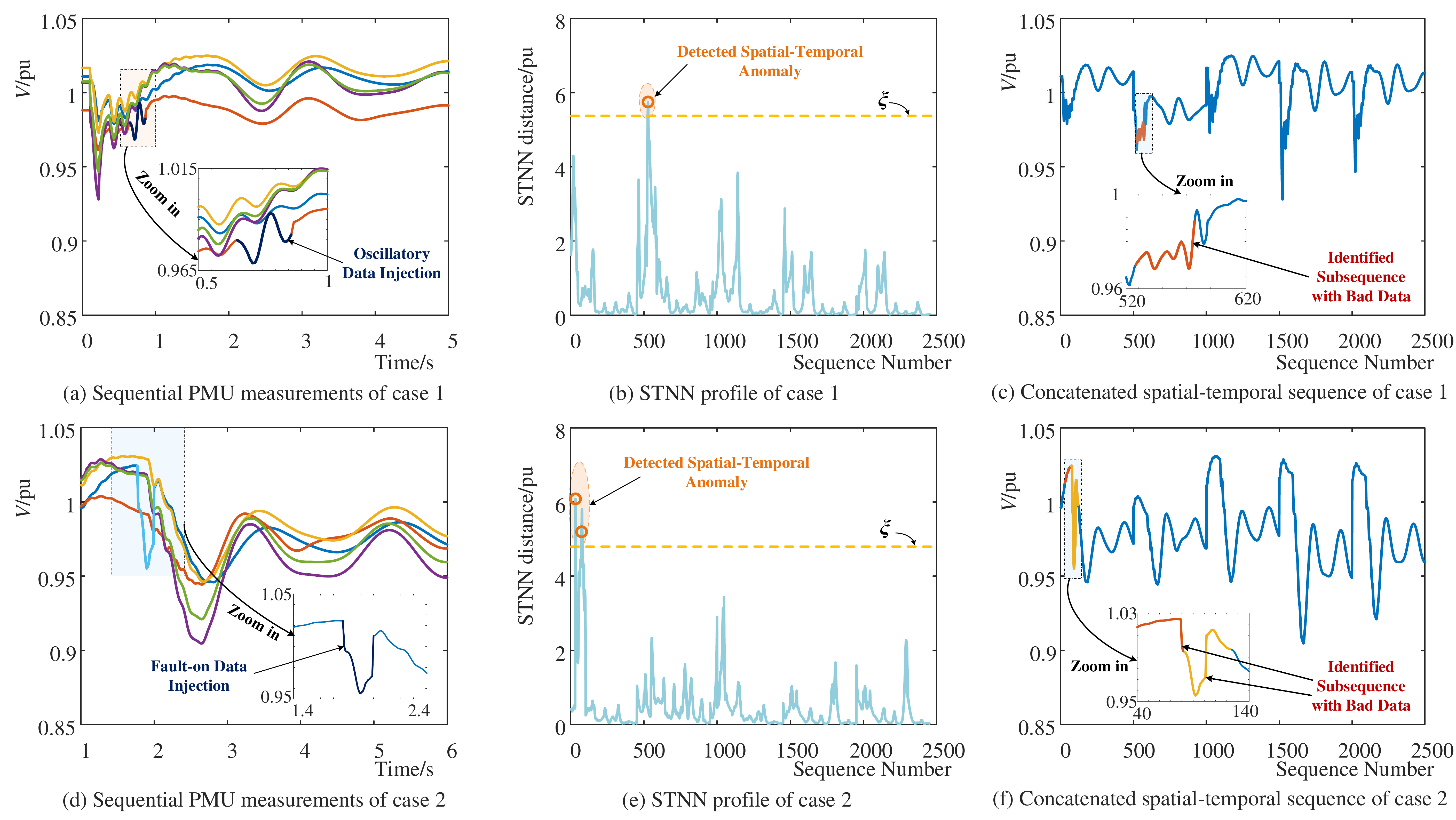}
	\caption{BPDD against false data injection. (case 1 $\rightarrow$ oscillatory data injection in transient process; case 2 $\rightarrow$ fault-on data injection in post-fault stage.)}
	\label{Fig4_scenario3}
\end{figure*}

For the second case, the fault-on response of a bus obtained from a historical event is injected into its current post-fault stage to mimic sequential transient events. Without additional attention to PMU measurements at other buses, it would be difficult for a single substation to identify this malicious data injection. By checking if the measurement matrix pertinently reflects the inherent spatial-temporal correlations, however, the proposed BPDD approach is able to precisely recognize this deceptive data segment.}

\subsection{Comprehensive BPDD Performances}{\color{black} For the purpose of comprehensively verifying the proposed approach's BPDD performances, all the PMU data matrices were fed into it for statistical tests. To demonstrate its superiority, it was compared with some representative data-driven BPDD methods in the literature, including SC \cite{yang2020bad}, LOF analysis \cite{wu2017online}, PCA \cite{mahapatra2016bad} and ANN \cite{mahapatra2017online}. As most of these methods \cite{yang2020bad, wu2017online, mahapatra2017online} require either a clean dataset or a labelled dataset for parameter setting or offline training, the 8000 PMU data matrices initially generated in Section \ref{section 4}-A were taken as an offline known dataset to fulfil this need. To make a fair comparison, another group of 4000 PMU data matrices simulating various online operating conditions and contingencies different from those considered in Section \ref{section 4}-A were produced and gathered as an unseen set for online tests. Four metrics, i.e., misdetection rate, false alarm rate, precision and accuracy \cite{zhu2017imbalance, han2011data}, were calculated to evaluate each method's BPDD performances:
\begin{equation}
Mis = {n_{fn}}/{n_{\text{all}}} \times 100\%~~~~~~
\label{eq_Mis}
\end{equation}
\begin{equation}
Fal = {n_{fa}}/{n_{\text{all}}} \times 100\%~~~~~~
\label{eq_Fal}
\end{equation} 
\begin{equation}
Prec = n_{ta}/(n_{ta} + n_{fa})  \times 100\%
\label{eq_Pre}
\end{equation}
\begin{equation}
Acc = \left( {n_{\text{all}}} - n_{fn} - n_{fa} \right)\left /{n_{\text{all}}} \times 100\% \right.
\label{eq_Acc}
\end{equation}  
where $n_{\text{all}}$ is the total number of instances for test; $n_{fn}$ is the number of falsely dismissed anomalous instances (with bad data); $n_{fa}$ is the number of normal instances (with no bad data) falsely identified as anomalies; $n_{ta}$ is the number anomalous instances truly detected as anomalies. Based on these metrics, all the methods carried out BPDD on the 8000 offline known cases and the 4000 online unseen cases. Relevant parameters and thresholds in these comparative methods were carefully tuned by using the offline case set to help them achieve their optimal performances. The statistical BPDD results of all the methods are summarized in Tables~\ref{Tab_BPDD_Training} and \ref{Tab_BPDD_Testing}.

\begin{table}[t]
	\color{black} 
	\centering
	\caption{BPDD Performances on Nordic System (Offline Known Cases)}
	\renewcommand\arraystretch{1.05}
	\newcommand{\tabincell}[2]{\begin{tabular}{@{}#1@{}}#2\end{tabular}}	
	\begin{tabularx}{0.485\textwidth}{m{1.2cm}<{\centering} m{1.3cm}<{\centering} m{1.3cm}<{\centering} m{1.3cm}<{\centering} m{1.3cm}<{\centering}}	
		\hline\hline 
		Method & $Mis$/\%   & $Fal$/\%  & $Pre$/\% & $Acc$/\%\\		
		\hline 
		Proposed & 0.99  &3.80 & 93.95 &95.21\\
		SC \cite{yang2020bad} & 1.35 & 1.94  & 96.80 & 96.71\\
		LOF \cite{wu2017online} & 3.25 & 1.41 & 97.57 & 95.34\\
		PCA \cite{mahapatra2016bad} & 4.05 & 6.63 & 89.41 & 89.32\\
		ANN \cite{mahapatra2017online} & 1.31 & 2.38 & 96.11 & 96.31\\
		\hline\hline
	\end{tabularx} \label{Tab_BPDD_Training}
		
\end{table}

\begin{table}[t]
	\color{black} 
	\centering
	\caption{BPDD Performances on Nordic System (Online Unseen Cases)}
	\renewcommand\arraystretch{1.05}
	\newcommand{\tabincell}[2]{\begin{tabular}{@{}#1@{}}#2\end{tabular}}	
	\begin{tabularx}{0.485\textwidth}{m{1.2cm}<{\centering} m{1.3cm}<{\centering} m{1.3cm}<{\centering} m{1.3cm}<{\centering} m{1.3cm}<{\centering}}	
		\hline\hline 
		Method & $Mis$/\%   & $Fal$/\%  & $Pre$/\% & $Acc$/\%\\	
		\hline 
		Proposed & 1.10  &3.67 & 94.13 &95.23\\
		SC\cite{yang2020bad} & 3.85 & 7.20 & 88.63 & 88.95\\
		LOF \cite{wu2017online} & 2.93 & 4.08 & 93.34 & 92.99\\
		PCA\cite{mahapatra2016bad} & 4.23 & 5.97 & 90.32 & 89.80\\
		ANN\cite{mahapatra2017online} & 2.78 & 4.10 & 93.31 & 93.12\\
		\hline\hline
	\end{tabularx} \label{Tab_BPDD_Testing}
		
\end{table}

As can be observed, most of the methods except the PCA alternative achieve relatively high BPDD accuracies ($>$ 90\%) on the offline cases. In particular, the SC and ANN methods with iterative offline training have the highest accuracies on the known set. The proposed method and the LOF analysis alternative also achieve satisfactory BPDD accuracies, though being a bit lower than the former two. However, as presented in Table~\ref{Tab_BPDD_Testing}, when faced with online unseen cases, the SC and ANN methods fail to generalize their performances to such unfamiliar conditions, with the SC method performing even worse than the PCA alternative. For the LOF analysis method, its BPDD performance is also degraded. In fact, although it has no offline training procedure, its parameter setting is affected by the distribution of clean data. Due to the deviation of the data distribution in unforeseen cases, the parameters obtained from known cases may not be optimal any more. 

In contrast, the proposed approach without iterative learning or data distribution related parameter setting is able to maintain it high performance. Clearly, it outperforms the remaining methods on the unseen cases, with the overall BPDD accuracy being 2.1$\sim$6.3\% higher. Hence, the proposed approach can adapt well to practical online monitoring environments where unforeseen system operation changes and events occur constantly. In addition, it is noticed that the proposed approach achieves the lowest misdetection rate among all the methods. This is a desirable merit for the BPDD task, because misdetection of bad data may be difficult to be remedied, while falsely alarmed normal data can be approximately recovered by appropriate data correction tools. Considering these advantages, the proposed approach would be preferred in practice.}

\subsection{Computational Efficiency of BPDD} The computational efficiency of the BPDD approach was tested here by recording its computation time in each case. All the tests were carried out using a PC configured with a 3.60-GHz$*$8 Intel Core i7-7700 CPU and 32.0 GB RAM. For comparative study, a brute-force method without adoption of the fast STNN discovery strategy (see Section III-B) is conducted for BPDD as well. {\color{black} The time consumptions of the two methods in the known and unseen cases are summarized in Table~\ref{Tab_computation_time}.} Evidently, after the adoption of the fast STNN discovery strategy, the computational efficiency of BPDD is dramatically improved by more than 10 times. Concretely, it costs the proposed approach less than 0.35 s to complete BPDD for a OTW of 5 s. In practical onling monitoring, highly efficient streaming BPDD can be performed by continuously sliding the OTW with a time step of 0.4$\sim$0.5 s.       
\begin{table}[t]
	\color{black}
	\centering
	\caption{BPDD Computational Cost}
	\renewcommand\arraystretch{1.05}
	\newcommand{\tabincell}[2]{\begin{tabular}{@{}#1@{}}#2\end{tabular}}	
	\begin{tabularx}{0.485\textwidth}{m{1.6cm}<{\centering} m{3cm}<{\centering} m{3cm}<{\centering}}	
		\hline\hline 
		Method & Average execution time (8000 known cases)   & Average execution time (4000 unseen cases) \\	
		\hline 
		Proposed & 0.324 s &0.326 s\\
		Brute-force & 3.769 s   & 3.770 s\\
		\hline\hline
	\end{tabularx} \label{Tab_computation_time}

\justifying{\noindent {\small {
{\it {\textbf{Remark 1}}}:For each case, the computation procedure was repeated 10 times to yield a stationary estimation of its execution time.}}}
\end{table}

\section{Experimental Test in Practical System} \label{section 5}
\begin{figure*}[t]
	\centering
	\includegraphics[width=0.8\linewidth]{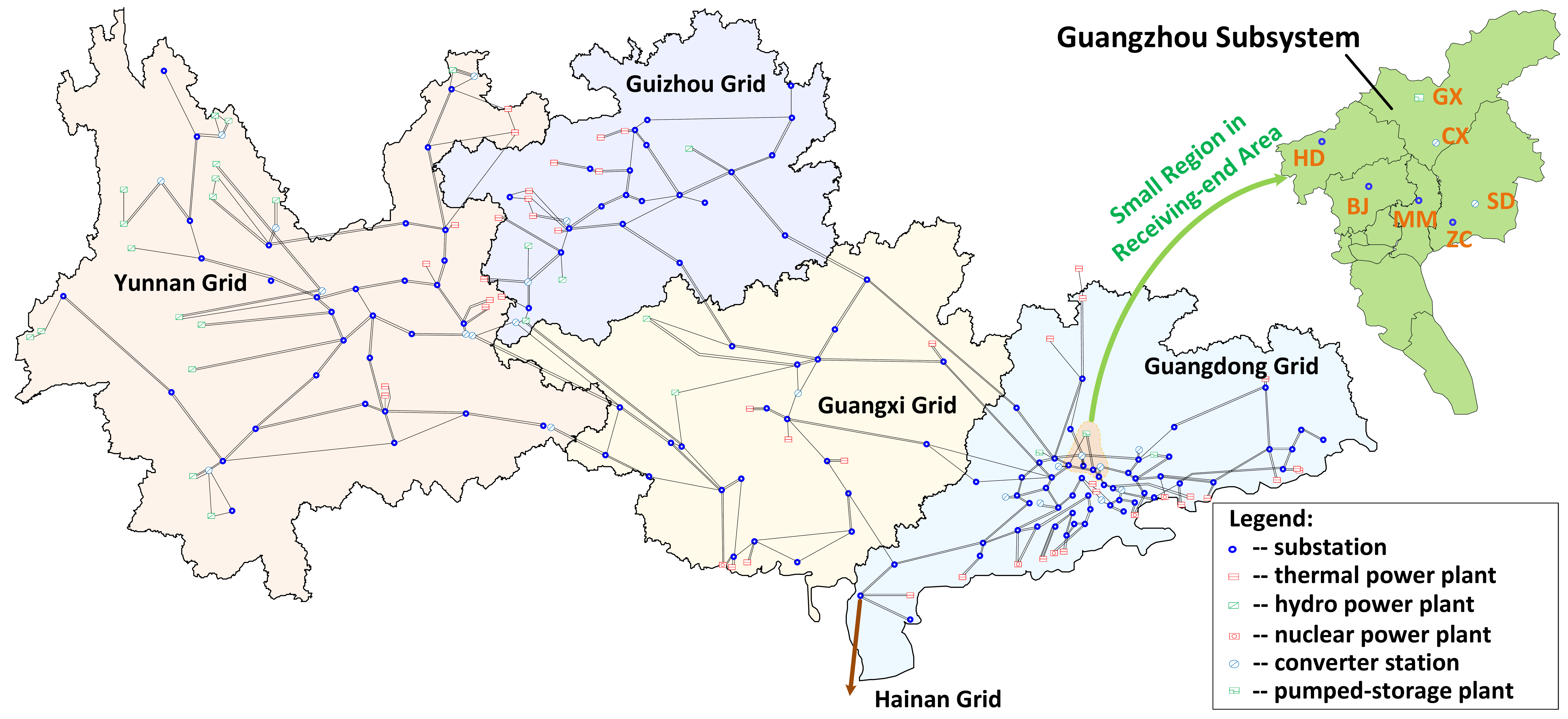}
	\caption{Structure of China Southern Power Grid \cite{zhu2020timeseries}.}
	\label{Fig5_CSG}
\end{figure*}

{\color{black} With field PMU data collected from the real-world CSG in two months of 2018,} the proposed approach was further tested in practical contexts to show its applicability and scalability. In particular, synchronous voltage measurement matrices with a OTW of 16 s were acquired from CSG for case study here. The PMU sampling rate was set to 25 Hz in CSG. Taking a region in Guangdong Province with seven 500-kV buses for example, i.e., the Guangzhou subsystem depicted in Fig.~\ref{Fig5_CSG} \cite{zhu2020timeseries}, the seven buses' voltage profiles were collected for BPDD tests. {\color{black} Similar to the tests on the Nordic system, two groups of datasets, i.e., an offline known case set and an online unseen case set, were generated from the two months' field PMU data, respectively. For each set, in addition to the actual bad data conditions, synthetic bad data scenarios (including data spike, repeated data and false data injection) were considered to augment its diversity. By doing so, 12000 known cases and 12000 unseen cases were prepared for BPDD tests.}      

{\color{black}
\subsection{Practical BPDD Scheme}} Unlike the simulation based PMU measurements in test systems, as there may exist multiple channels to measure bus voltages at a single substation (bus), multiple voltage TS can be obtained for the same bus in the bulk system. Considering this practical situation, a two-layer BPDD scheme was designed, {\color{black} as illustrated in Fig.~\ref{Fig6_bpdd_scheme}. With a strategic combination of practical engineering BPDD rules and the proposed BPDD approach, the BPDD scheme is detailed in the following. 
 
\begin{figure}[t] 
	\color{black}
	\centering
	\includegraphics[width=1.0\linewidth]{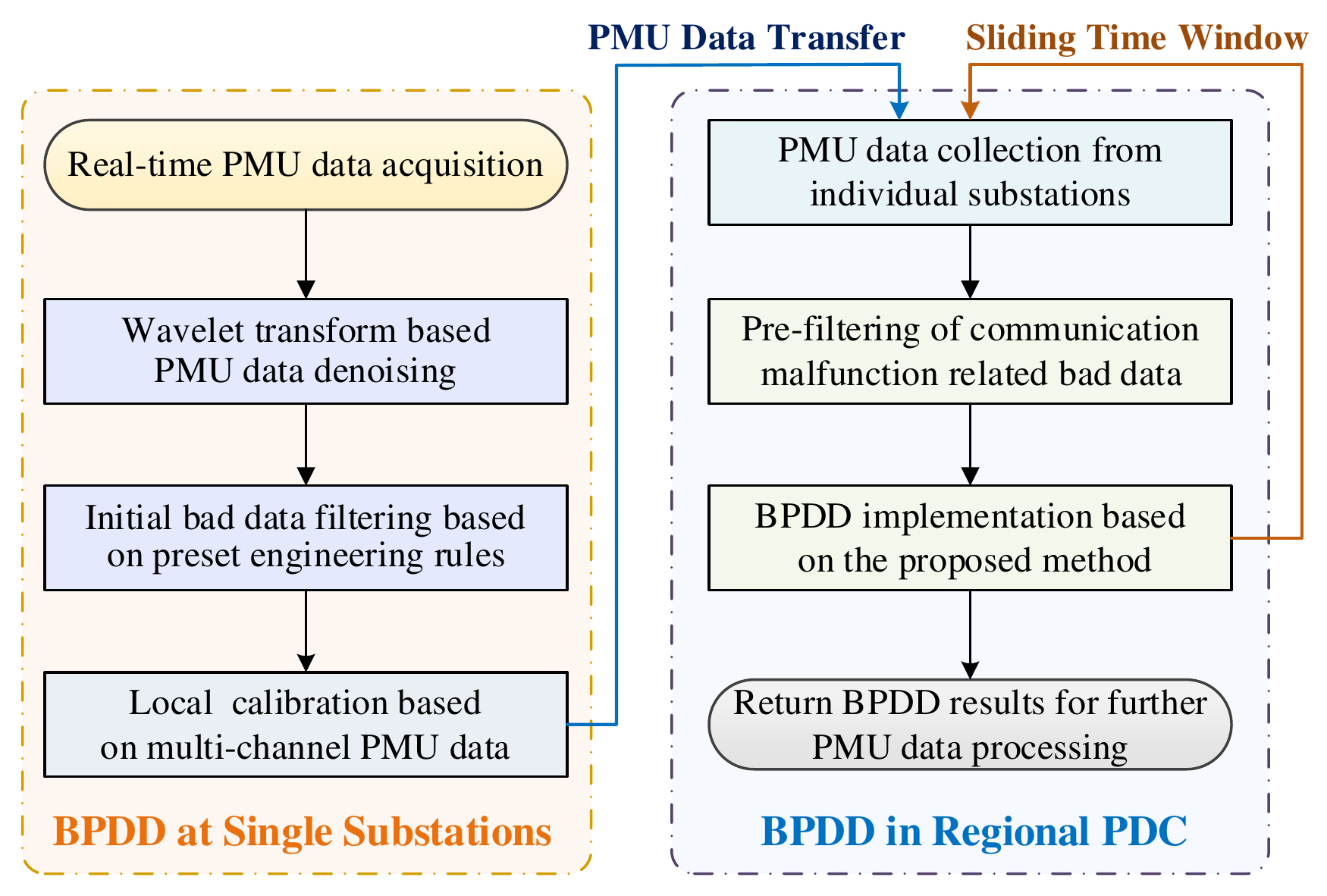}
	\caption{Two-layer BPDD scheme designed for CSG.}
	\label{Fig6_bpdd_scheme}
\end{figure} 
 
\subsubsection{BPDD at Single Substations}~At a single substation, PMU data are assumed to be acquired in real time. WT-based signal processing is first performed to filter out possible noises in the raw PMU measurements. Then, based on engineering rules \cite{martin2014synchrophasor} preset by system practitioners, obvious bad PMU measurements such as missing data and data points with impractical/invalid values are flagged. If multiple channels of PMU data are available at the substation, cross calibration is carried out locally by comparing multi-channel PMU measurement profiles to further detect latent anomalous PMU data. In fact, as the comparison of multi-channel PMU data has already implicitly applied the idea of spatial-temporal dissimilarity, the proposed BPDD approach is not repetitively deployed at substation levels. Those identified bad data can be further fixed with efficient interpolation techniques such as linear/derivative-based PMU data estimation \cite{pourramezan2017design}. Afterwards, the processed local PMU data are uploaded to the corresponding regional PDC for further concentration.  

\subsubsection{BPDD in Regional PDC}~The above BPDD procedure may bypass some bad data if only a single channel of data is available at a given substation. Besides, it would be incompetent in conditions where multi-channel measurements encounter consistent data issues (e.g., temporary un-updated data) or false data injection caused by cyber attacks within the whole substation. For these cases, the approach proposed in this paper can act as the second defensive line against bad PMU data in regional PDCs. Note that, for a certain PDC gathering PMU data from regional substations, as imperfect data transfer in communication networks may cause visible bad data such as data dropout and delayed arrival, these anomalies are first filtered out. Then, with an OTW, the proposed BPDD approach can be efficiently carried out to identify latent bad data. After the completion of its computation, the BPDD results are issued for further PMU data processing, e.g., bad data correction. Meanwhile, the OTW would slide along with time to continuously process PMU data streams.  
 
As the proposed approach mainly involves BPDD in the regional PDC layer, the following  tests will focus on BPDD in the regional PDC of Guangzhou.
}   

\begin{figure*}[t] 
	\color{black}
	\centering
	\includegraphics[width=1.0\linewidth]{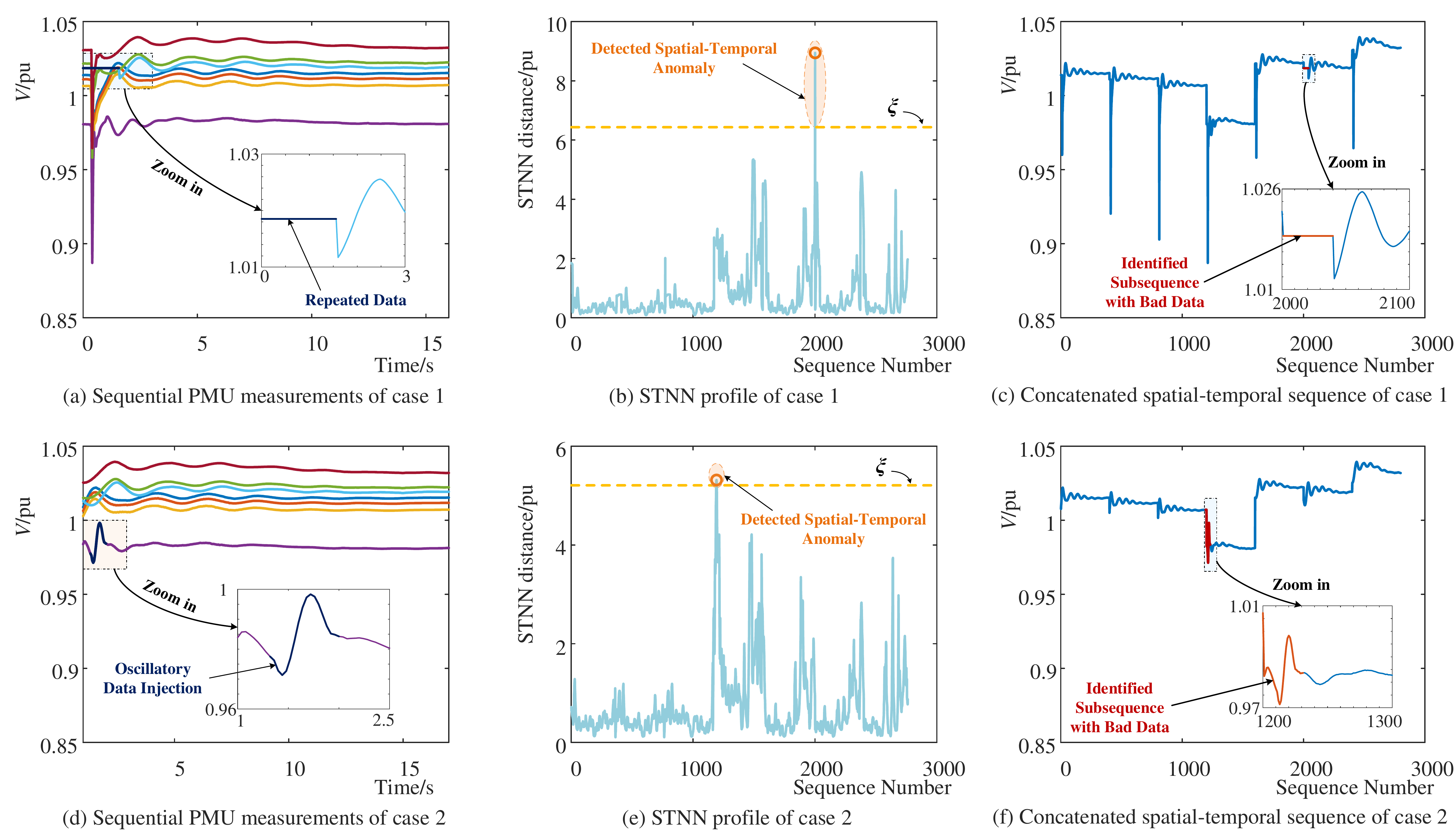}
	\caption{Practical BPDD example. (case 1 $\rightarrow$ repeated data in transient process; case 2 $\rightarrow$ false data injection in post-fault stage.)}
	\label{Fig7_CSG_bpdd}
\end{figure*}

{\color{black}
\subsection{Illustration of BPDD}
Without loss of generality, two cases were randomly chosen from the unseen dataset to illustrate BPDD in the regional PDC, as depicted in Fig.~\ref{Fig7_CSG_bpdd}. The first case comes from an actual scenario, where one of the seven substations undergoes un-updated measurements for 1.6 s, which ranges from the pre-fault stage to the post-fault stage. As shown in Fig.~\ref{Fig7_CSG_bpdd}(a), for the single bus voltage profile, it looks like a small disturbance caused by a sudden load increase rather than a drastic transient event. With only one channel of voltage measurements available at this substation, the first-layer BPDD at the substation level fails to identify this bad segment. Based on the contemporary measurements collected from other buses, such an apparently misleading case is accurately identified by the second-layer BPDD in the PDC [see Fig.~\ref{Fig7_CSG_bpdd}(b)$\sim$(c)].

The second case corresponds to a synthetic scenario with false data injection at a single substation. As illustrated in Fig.~\ref{Fig7_CSG_bpdd}(d), a segment of oscillatory measurements obtained from historical PMU records are injected into the post-fault process. While this deceptive case may not be easily recognized locally, it is correctly detected by the proposed method configured in the regional PDC, with a systematic exploration of the differences of oscillatory profiles between adjacent buses. Both of the two cases reveal that the proposed approach can help screen bad data effectively for the regional PDC.}         

\subsection{Comprehensive BPDD Performances}~{\color{black}To further evaluate the proposed approach's overall BPDD performances, all the cases generated above were examined one by one. Following the previous settings of comparative study in Section \ref{section 4}-D, the comparative methods of SC \cite{yang2020bad}, LOF analysis \cite{wu2017online}, PCA \cite{mahapatra2016bad} and ANN \cite{mahapatra2017online} were realized for comparison again. To make a reasonable comparison, two-layer BPDD was also implemented for these methods, with each of them replacing the proposed approach in the second-layer BPDD. Taking all the known cases as the offline dataset for their parameter setting or iterative training, their BPDD performances in the second layer were tested on the 12000 online unseen cases, as summarized in Table~\ref{Tab_BPDD_Testing_CSG}. 

\begin{table}[t]
	\color{black} 
	\centering
	\caption{BPDD Performances on CSG (Online Unseen Cases)}
	\renewcommand\arraystretch{1.05}
	\newcommand{\tabincell}[2]{\begin{tabular}{@{}#1@{}}#2\end{tabular}}	
	\begin{tabularx}{0.485\textwidth}{m{1.2cm}<{\centering} m{1.3cm}<{\centering} m{1.3cm}<{\centering} m{1.3cm}<{\centering} m{1.3cm}<{\centering}}	
		\hline\hline 
		Method & $Mis$/\%   & $Fal$/\%  & $Pre$/\% & $Acc$/\%\\	
		\hline 
		Proposed & 0.55  &3.78 & 94.02 &95.67\\
		SC\cite{yang2020bad} & 2.73 & 8.06 & 87.66  & 89.21\\
		LOF \cite{wu2017online} & 2.85 & 4.10 & 93.31  & 93.05\\
		PCA\cite{mahapatra2016bad} & 3.89 & 6.01 & 90.33  & 90.10\\
		ANN\cite{mahapatra2017online} & 1.99 & 5.02 & 92.04 & 92.99\\
		\hline\hline
	\end{tabularx} \label{Tab_BPDD_Testing_CSG}
\end{table}

Obviously, the proposed approach achieves the best BPDD performance, with the overall accuracy remaining above 95.5\%. In fact, analogous to the comparative results in Section \ref{section 4}-D, this is because other methods with a heavy reliance on a known dataset for parameter setting or offline training cannot adapt well to the unseen dataset. Meanwhile, it is found that the proposed approach has the lowest risk of falsely dismissing bad data. Among the 12000 unseen instances, only 66 ones with bad data are wrongly bypassed. Given such a high reliability, the proposed BPDD approach would significantly contribute to low-quality data filtering in regional PDCs of pratcical power grids.}

{\color{black}
\section{Further Discussion} \label{section 6}
The above test results have exhibited the efficacy of the proposed approach for BPDD in regional PDCs. It is able to efficiently identify anomalous PMU measurements bypassed by individual substation-layer detections. These identified anomalies as well as other low-quality data such as missing values caused by communication issues should be further corrected to improve the PMU data quality eventually. In this respect, the proposed approach can be employed as a standard preprocessing tool and integrated with existing data correction techniques to build a comprehensive PMU data cleansing solution. For instance, by combining it with the adaptive PMU data estimator proposed in \cite{pourramezan2017design}, regional PDCs may not only fix missing data points, but also correct those anomalies detected by itself. 
Besides, if the inherent spatial-temporal correlations within regional PMU data are further explored via well-designed ML techniques for intelligent matrix data interpolation, new PMU data correction methods based on spatial-temporal data analytics can also be developed. This can be taken as one of the future directions for extensive study.

Moreover, it should be noted that the proposed BPDD approach can work effectively under most normal conditions where multiple substations do not encounter the same anomalies at the same time instant. It is because individual sensing and communication devices within different substations rarely experience the same data problem at the same time. However, for premeditated cyber attacks at multiple substations, this may not hold, as consistent false data could be concurrently injected into multiple substations. While how to devise a systematic defensive scheme against such highly deceptive data injection attacks is not the main focus in this paper, future research efforts may be devoted to tackling this challenging issue. For instance, more abundant information such as $k$ ($k > $ 1) nearest spatial-temporal neighbors can be extracted to characterize anomalous behaviors more comprehensively. In addition, more intelligent ML-based rules rather than the simple threshold-based rule in \eqref{threshold} can be designed to detect various complicated data injection scenarios more accurately.
} 

\section{Conclusion} \label{section 7}
{\color{black} Based on the inherent spatial-temporal correlations during power system dynamics, this paper develops a model-free TS data-driven approach for online BPDD in regional PDCs.} With no need for labeling bad PMU data in advance for offline learning, it performs unsupervised BPDD in a a highly efficient way. Specifically, following the idea that spatial-temporal anomalies are significantly different from their STNNs, sequential BPDD is carried out by performing fast STNN discovery and checking TS subsequences with abnormal STNN distance values. With no requirement on iterative learning, it gets rid of time-consuming offline learning, being suitable for handling online PMU data streams. Numerical test results on the Nordic test system show that the proposed approach achieves excellent performances in various bad data scenarios. Further tests with field PMU data in CSG demonstrate the scalability of the BPDD approach in practical contexts. 

{\color{black} In relevant future work, how to further correct those bad PMU data identified by the BPDD approach remains to be investigated. Research efforts may also be devoted to working out a more powerful bad data detection scheme against more challenging false data injection conditions.}


%
\bibliographystyle{IEEEtran}
\bibliography{ieee_paper}
\ifCLASSOPTIONcaptionsoff
  \newpage
\fi

\end{document}